\documentstyle[aps,prb,eqsecnum,multicol,epsf]{revtex}
\begin{document}
\input epsf
\bibliographystyle{prsty}
\draft
\preprint{NSF-ITP}
\title{Transport of Surface States in the Bulk Quantum Hall Effect}
\author{Sora Cho}
\address{Department of Physics, University of California,
        Santa Barbara, California 93106}
\author{Leon Balents}
\address{Institute for Theoretical Physics,
        University of California, Santa Barbara, California 93106}
\author{Matthew P.A. Fisher}
\address{Institute for Theoretical Physics,
        University of California, Santa Barbara, California 93106\\
        and Department of Physics, University of California,
        Santa Barbara, California 93106}
\date{\today}
\maketitle

%%%%%%%%%%%%%%%%%%%%%%%%%%%%%%%%%%%%%%%%%%%%%%%%%%%%%%%%%%%%%%%%%%%%%%%
\begin{abstract}
  The two-dimensional surface of a coupled multilayer integer quantum
  Hall system consists of an anisotropic chiral metal.  This unusual
  metal is characterized by ballistic motion transverse and diffusive
  motion parallel ($\hat{z}$) to the magnetic field.  Employing a
  network model, we calculate numerically the phase coherent
  two-terminal z-axis conductance and its mesoscopic fluctuations.
  Quasi-1d localization effects are evident in the limit of many
  layers.  We consider the role of inelastic de-phasing effects in
  modifying the transport of the chiral surface sheath, discussing
  their importance in the recent experiments of Druist
  {\sl et~al.}\cite{druist97}\ 

% The surface state of a coupled multilayer quantum Hall system at
% a plateau $\nu=1$ is considered to be a very anisotropic chiral metal, 
% which shows ballistic motion
% transverse and diffusive motion parallel to the magnetic field.
% We calculate the averaged two-terminal conductance of the
% system and its fluctuations.  We considered in the limit of
% $L^2 \gg DC$ so called one-dimensional limit where
% D is a diffusion constant and show the universal conductance
% fluctuation.
\end{abstract}
%%%%%%%%%%%%%%%%%%%%%%%%%%%%%%%%%%%%%%%%%%%%%%%%%%%%%%%%%%%%%%%%%%%%%%%

\begin{multicols}{2}
%%%%%%%%%%%%%%%%%%%%%%%%%%%%%%%%%%%%%%%%%%%%%%%%%%%%%%%%%%%%%%%%%%%%%%%
\section{Introduction}
%%%%%%%%%%%%%%%%%%%%%%%%%%%%%%%%%%%%%%%%%%%%%%%%%%%%%%%%%%%%%%%%%%%%%%%

In a two-dimensional incompressible quantized Hall state, the low
energy excitations are confined to the edge of the sample.  These edge
states provide a simple way to understand transport in both integer
and fractional quantum Hall systems.\cite{review} For the integer quantum Hall
effect with one filled Landau level, there is a single edge mode,
describable in terms of a free chiral Fermion.  Edge states in the
FQHE are believed to be (chiral) Luttinger liquids, and have been
probed via tunneling spectroscopy in several recent experiments.\cite{FQHE}

In recent years there has been much interest in multi-layer quantum
Hall systems.  In double layer systems the layer index plays the role
of a pseudo-spin, and these systems have revealed a number of new
surprises.  In the opposite extreme with many layers, the samples
become three-dimensional, and a number of new features are expected.
In such bulk samples with interlayer tunneling smaller than the Landau
level spacing, the (integer) quantized Hall effect in each layer
should survive, and the sample exhibit a 3d quantum Hall phase.
Chalker and Dohmen\cite{chalker95} have recently discussed the phase diagram in such a
system, in a model of non-interacting electrons with
disorder. In the absence of disorder, the Landau
levels will be broadened into bands in the presence of interlayer
tunneling, $t$.  Disorder further broadens these bands.  Near the band
centers a diffusing 3d metallic state is expected.  In the tails of
the Landau bands, the bulk states are localized,  but current
carrying edge states nevertheless lead to a quantum Hall
effect.  For one full Landau level, each layer has a single chiral
free Fermion edge state, which together comprise a 2d sub-system - a
chiral surface sheath.\cite{chalker95,balents96}  This surface phase
forms a novel 2d chiral metal system, which has been analyzed
theoretically by a number of authors.
\cite{mathur96,balents96b,yu96,gruzberg97b,kim96,gruzberg97}  In the
presence of impurity scattering, the transport is predicted to be very
anisotropic, with ballistic in-plane motion and diffusive motion
parallel to the magnetic field.  Vertical transport in such a multilayer
sample was first investigated experimentally in Ref.~\onlinecite{stormer86}, and has
recently been revisited by Druist {\sl et~al.}\cite{druist97}\  The latter
experiment provides striking evidence of the novel behavior
characteristic of the chiral metal.

\begin{figure}
\epsfxsize=2.0in
\centerline{\epsfbox{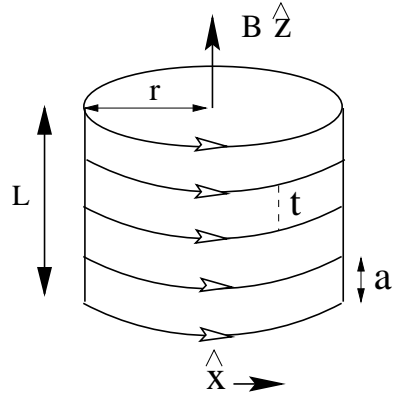}}
\begin{minipage}[t]{8.1cm}
\caption{Geometry of a multilayer quantum Hall system with
an interlayer hopping amplitude $t$ allowing vertical  
transport.  With $N$ layers the system has height
$L=Na$, and a circumference $C=2\pi r$.
\label{system}}
\end{minipage}
\end{figure}

Most of the theoretical work on the chiral metal phase has focused on
the mesoscopic regime, with the sample assumed smaller than the phase
breaking lengths.  The predicted behavior for such phase coherent
transport is very rich, with three different possible regimes (see
Fig.~\ref{regime}) connected by universal
crossovers.\cite{balents96,mathur96,balents96b,gruzberg97b}\ 

The purpose of this paper is two-fold.  First, we revisit the phase
coherent regime and study in detail the conductance and its
fluctuations.  Performing numerical transfer matrix calculations on a
directed network model for the chiral metal enables us to extract the
conductance and its fluctuations in the various regimes, and compare
directly with earlier analytic approaches.  We then address the
important role of phase breaking processes, which have been ignored in
earlier theoretical discussions.

The paper is organized as follows.  In Section II we briefly review the 
existing theoretical predictions for the phase coherent transport.
In Section III we describe the network model, and extract
numerically the phase coherent conductance in the various regimes.
Section IV is devoted to a discussion of de-phasing processes,
and Section V to prospects and conclusions.

%%%%%%%%%%%%%%%%%%%%%%%%%%%%%%%%%%%%%%%%%%%%%%%%%%%%%%%%%%%%%%%%%%%%%%%
\section{Phase Coherent Regime}
%%%%%%%%%%%%%%%%%%%%%%%%%%%%%%%%%%%%%%%%%%%%%%%%%%%%%%%%%%%%%%%%%%%%%%%

For one full Landau level, there is a single
free chiral Fermion edge mode in each layer, as depicted
in Fig.~\ref{system}.  In the presence of an interlayer tunneling
amplitude, $t$ (assumed much smaller than $\hbar \omega_c$),
these chiral edge modes disperse along the z-axis, and form
one-half of an open 2d Fermi surface.
Impurities causes electrons to scatter
about the Fermi surface, as in any dirty metal.
Due to the chiral nature, the in-plane motion
remains ballistic with velocity $v$, even in the presence of impurities.
However, the (inter-layer) motion parallel to the field becomes
diffusive, with diffusion constant $D$.
Easier to measure than the ballistic velocity or diffusion constant
is the z-axis (2d sheet) conductivity, $\sigma_{zz}$,
related to $v$ and $D$ via an Einstein relation:\cite{balents96}
\begin{equation}
 \sigma_{zz}=e^2\rho D = {D\over va}\cdot{e^2\over h}
\end{equation}
where $a$ is the inter-layer (lattice) spacing
and $\rho=1/hva$ the density of states.
It will be convenient to introduce a 
dimensionless z-axis conductivity
via $\sigma_{zz} = (e^2/h) \sigma$.

For a mesoscopic sample with finite circumference, $C$, and number of
layers, $N=L/a$, there are several important time scales.  For
ballistic motion with velocity $v$, an electron circumnavigates the
sample in a time $\tau_c = C/v$.  In a time $\tau_L = L^2/D$ an
electron will diffuse from the bottom to the top of the sample.  The
transport will be phase coherent provided the de-phasing time,
$\tau_{\phi}$, is much longer than both $\tau_c$ and $\tau_L$.  In
principle, this mesoscopic regime exists for any sample at
sufficiently low temperatures, since the de-phasing time diverges as $T
\rightarrow 0$ ($\tau_\phi \sim \hbar/k_{\rm B} T$ in the quasi-1d
limit of interest).  Here we focus on the fully coherent regime,
returning to de-phasing effects in Section V.

\begin{figure}
\epsfxsize=3in
\centerline{\epsfbox{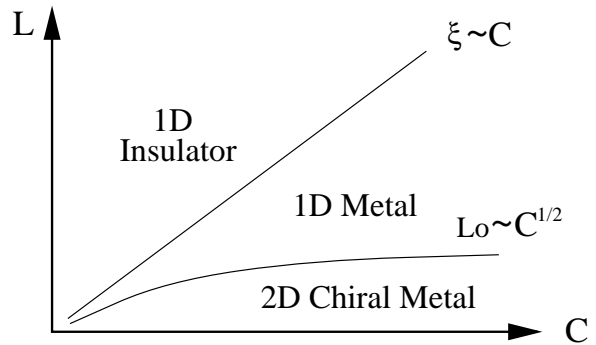}}
\begin{minipage}[t]{8.1cm}
\caption{Three different regimes of the phase coherent transport
of the surface sheath of a sample with
height $L$ and circumference $C$.
Both $L_0$, the typical distance an electron diffuses along
the z-axis upon circling the sample once, and $\xi$,
the 1d localization length, can be deduced from the measurable
z-axis conductivity and geometrical parameters, as discussed in the text.
\label{regime}}
\end{minipage}
\end{figure}

For a sample with finite circumference, $C$, there
are two important length scales along the z-axis, which
demarcate the boundaries between three regimes (see Fig.~\ref{regime}).
\cite{balents96,mathur96,gruzberg97b}
Upon circumnavigating the sample once, an electron
will diffuse along the z-axis a distance $L_0 = \sqrt{D\tau_c}$,
which can be expressed in terms of the measurable 
z-axis conductivity, $\sigma$, as,
\begin{equation}
 L_0=(a\sigma C)^{1/2}.
\label{L0def}
\end{equation}
For finite $C$ with $L \rightarrow \infty$
the system is one-dimensional, and localization along the z-axis is expected.
The (typical) localization length, $\xi$, for
such a quasi-1d system is proportional to
the (dimensionless) 1d conductivity, $\xi \sim \sigma_{1d}$,
which can be written,
\begin{equation}
\xi = 2 \sigma C  .
\label{xidef}
\end{equation}
Thus both $L_0$ and $\xi$ depend only on geometrical parameters, and
the measurable z-axis conductivity, $\sigma$.
Notice that $(\xi/a) = 2(L_0/a)^2$,
so that provided $L_0 \gg a$ one has $\xi \gg  L_0$.  

As the height $L$ of the sample varies,
three regimes are possible (see Fig.~\ref{regime}).
For $L < L_0 \ll  \xi$, an electron typically
diffuses from the bottom to the top of the sample
before circumnavigating the sample once.
In this {\sl 2d chiral metal} regime,
an electron suffers de-phasing in the leads before
circling the sample.
For $L_0 \ll  L \ll  \xi$, the electron circles the sample many times,
and phase coherent processes around the sample are important.  The
system behaves like a phase coherent {\sl quasi-1d metal}.  Finally, for $L \gg  \xi$ 1d localization effects
dominate,
and the system is a {\sl 1d} (localized) {\sl insulator}.

The predicted behavior for the phase coherent z-axis conductance
and it's mesoscopic fluctuations depends sensitively on which regime 
the system is in.  Consider first the (dimensionless)
mean two-terminal conductance
along the z-axis,
$\overline{G}$, where the overbar denotes an average over
disorder realizations.
In both the 2d chiral metal and the 1d metal regimes,
ohmic behavior is predicted with\cite{gruzberg97b},
\begin{equation}
 \overline{G}={C\over L}\sigma  + O(L/\xi) .
\end{equation}
The usual ``weak localization" corrections
which are of order $(L/\xi)^0$ are absent due to the breaking
of time reversal invariance.
In the 1d insulating regime strong localization is operative, 
and the mean conductance is predicted
to fall off exponentially with a universal 
form (for $L\gg  \xi$),\cite{mirlin94}
\begin{equation}
\overline{G} =2(\pi \xi/2L)^{3\over 2} \exp(-L/2\xi).
\end{equation}

Conductance fluctuations are also of interest, which can be characterized
by the variance, $\overline{\delta G^2}$, where
$\delta G = G - \overline{G}$.  In the 2d chiral metal and 1d metal regimes,
Gruzberg {\sl et~al.}\cite{gruzberg97} have shown that the variance can be written,
\begin{equation}
\overline{\delta G^2} = \Phi(L/L_0) + O(L/\xi),
\end{equation}
where $\Phi(X)$ is a universal scaling function which smoothly connects
the two regimes.  Deep within the 1d metal regime
the variance approaches a universal number well known for
quasi-1d metals: $\Phi(L/L_0 \rightarrow
\infty) = 1/15$.  In the 2d chiral metal regime, $\Phi(L/L_0) \sim (L_0/L)^2$
for $L/L_0$ small.  The conductance fluctuations are large in this limit,
since the sample effectively breaks up into $\tau_c/\tau_L = (L_0/L)^2$
incoherent regions which add independently to the conductance
and it's fluctuations.  Gruzberg {\sl et~al.}\cite{gruzberg97} 
have obtained the full universal
scaling function, $\Phi(X)$, which interpolates between these two limits.
In the 1d localized regime, the conductance is expected to be very broadly
distributed, with an approximate log-normal distribution.

%Balents and Fisher\cite{balents96} have studied conductance
%for a continuous model with
%a small interlayer matrix element t with Hamiltonian, 
%\begin{eqnarray}
% H & = & \sum_{i}\int dx \psi^\dagger_ii\hbar v\partial_x\psi_i
%- t(\psi^\dagger_i\psi_{i+1} + \psi^\dagger_{i+1}\psi_i)\nonumber\\
% & = & {1\over L}\sum_{p_z}^{BZ}\int {dp_x\over 2\pi}
% (\hbar vp_x-2t\cos p_za)\psi^\dagger_p\psi_p
%\end{eqnarray}
%where $v$ is the edge velocity, $a$ is the distance between two layers,
% and $t$ is tunneling amplitude.
%Therefore, without any disorder, there exist a gapless state when
%\begin{equation}
%\hbar vp_x=2t\cos p_za.
%\end{equation}

%%%%%%%%%%%%%%%%%%%%%%%%%%%%%%%%%%%%%%%%%%%%%%%%%%%%%%%%%%%%%%%%%%%%%%%
\section{Numerics}
%%%%%%%%%%%%%%%%%%%%%%%%%%%%%%%%%%%%%%%%%%%%%%%%%%%%%%%%%%%%%%%%%%%%%%%

\subsection{Network model}

\begin{figure}
\epsfxsize=3.5in
\centerline{\epsfbox{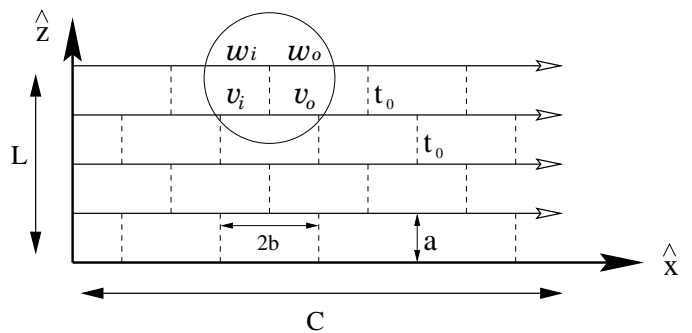}}
\begin{minipage}[t]{8.1cm}
\caption{Network model for the surface sheath.
Here $N=L/a = 4$ chiral edge modes are interconnected with
dimensionless tunneling $t_0$,
with periodic boundary conditions
taken in the $\hat{x}$ direction of circumference $C$.
The $z-$axis conductance is computed by employing
a transfer matrix acting in the $z-$direction.  
\label{fermion3}}
\end{minipage}
\end{figure}

Following Chalker and Dohmen,\cite{chalker95} we employ a simple network model
to study phase coherent transport of the surface sheath.
The network model consists of directed links carrying
electron current, connected via node parameters,
as depicted in Fig.~\ref{fermion3}.  All links carry current in the
$x-$direction, as appropriate for the chiral surface sheath.
Scattering at the nodes is characterized by a (real and
dimensionless) transmission amplitude, $t_0$, for tunneling in the
$z-$direction between edge states in neighboring layers.
For a given node the $S-$ matrix relating incoming to outgoing
amplitudes is given explicitly by,
\begin{equation}  
\left( \begin{array}{c} w_{\rm out} \\
v_{\rm out} \end{array} \right) =
 \left( \begin{array}{cc}
    r_0& t_0\\ t_0 & -r_0
\end{array} \right)
\left( \begin{array}{c} w_{\rm in} \\ v_{\rm in}  \end{array} \right) ,
\end{equation}
with $t_0^2+r_0^2=1$.
By construction, this matrix conserves the current,
$|w_{\rm in}|^2+|v_{\rm in}|^2 =|w_{\rm out}|^2+|v_{\rm out}|^2$.
To model the disorder, the electrons are assumed to acquire
a random phase along each link connecting adjacent nodes,
taken for simplicity to be independent and uniformly distributed on the interval $[0,2\pi]$.  

Periodic boundary conditions are taken in the ballistic $x-$direction,
with the circumference $C= 2b N_c$, where $b$ is the length of
a single link in the $x-$direction and $N_c$ is the total number
of inter-layer tunneling nodes connecting adjacent edge modes 
(see Fig.~\ref{fermion3}).  
The network consists of $N$ edge modes, with spacing $a$
and a total ``height" of $L=Na$.

The conductance along the $z-$axis is obtained by computing
the transmission of electrons from the bottom to the top of
the sample.  Specifically, we use the two-terminal Landauer formula
to relate the (dimensionless) conductance $G$
to the transmission matrix {\bf t}:\cite{dfisher81}
\begin{equation}
 G={\rm tr}[{\bf t}^+{\bf t}].
\end{equation}
The matrix elements, $t_{ij}$, are the amplitudes for
an electron incident into channel (or node) $i$ on the bottom edge
to be transmitted into channel $j$ on the top edge.
Here $N_c$ is the number of channels.

The transmission matrix is computed numerically by iterating a transfer matrix
from the bottom to the top of the sample. 
This involves re-expressing each node in a form relating
the amplitudes in one edge mode to the amplitudes in the adjacent
edge mode:
\begin{equation}
 \left( \begin{array}{c} w_{\rm in} \\
w_{\rm out} \end{array} \right) =
 \left( \begin{array}{cc}
    r_0/t_0 & 1/t_0\\ 1/t_0 & r_0/t_0
\end{array} \right)
\left( \begin{array}{c} v_{\rm in} \\ v_{\rm out} \end{array} \right) .
\end{equation}
We study a range of system sizes
with the channel number $N_c=4,8,16,32$ and the layer number
$N=8,10,12,16$.  
Being interested in conductance fluctuations,
it is necessary to evaluate
the conductance exactly for each given disorder
realization.
The self averaging Lyapunov exponents for
a sample with $L \rightarrow \infty$
cannot be used to extract the sample to sample fluctuations
in a finite system.
This restriction imposes rather serious constraints on
the accessible system sizes.

\begin{figure}
\epsfxsize=1.5in
\centerline{\epsfbox{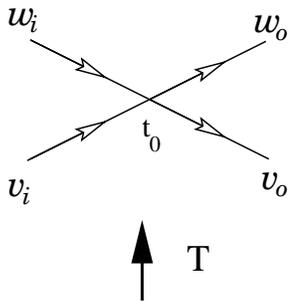}}
\begin{minipage}[t]{8.1cm}
\caption{The single node inside the circle in Fig.~3
%\ref{fermion3} 
(re-drawn in a more conventional way)
represented by the matrix in Eqn.~(3.3).
The transfer matrix progresses from bottom to top.
\label{node3d}}
\end{minipage}
\end{figure}

Since the microscopic parameters of the network model,
$t_0$ and $b$, are not experimentally meaningful quantities,
it is useful to relate them to a macroscopic observable,
namely the measurable $z-$axis sheet conductivity
of the surface sheath,
$\sigma$.  As shown by Chalker and Dohmen,\cite{chalker95} this is possible
for the network model, by summing the Feynman paths
analytically.  Specifically, consider paths that
connect the incident electrons on the bottom edge to the transmitted
electrons on the top edge.  For $C \rightarrow \infty$ these
paths do {\it not} fully circumnavigate the sample so
that the interference between paths wrapping around the sample
a different number of times - possible for finite $C$ - is
completely absent.  In the absence of such interference
the ensemble averaged conductance reduces to a
sum of classical probabilities:
Any two
paths which pass through a different sequence of directed links
will have a {\it random} relative phase,
so that the interference term will
vanish upon ensemble averaging.
To sum the classical probabilities of these non-winding paths 
we follow Chalker and Dohmen,\cite{chalker95} and consider the transmission
probability for an electron incident in one channel
(say $i$) to be transmitted through $N$ layers: $T_N = \sum_j |t_{ij}|^2$.
One can then express $T_{N+1}$ in terms of $T_N$ and
the single layer transmission probability, $T_1 = t_0^2$,
as a geometric sum:
\begin{equation}
 T_{N+1}=\sum_{n=0}^\infty T_N(R_1 R_N)^n T_1 ,
\end{equation}
where $R_N=1-T_N$ is the reflection probability off $N$ layers.
Carrying out this geometric sum, gives the recursion relation,
\begin{equation}
 {1\over T_{N+1}}= {1-T_1\over T_1} +{1\over T_N},
\end{equation}
which can be readily solved for $T_N$.  The average conductance
in the absence of interference between paths winding around the sample
is simply $\overline{G}_0 = N_c T_N$, with $N_c$ the number of channels.
It is given exactly by,
\begin{equation}
 \overline{G}_0 ={N_c\over N}{t_0^2\over 1-t_0^2(1-1/N)},
\label{G0}
\end{equation}
as obtained by Chalker and Dohmen\cite{chalker95} for $1/N \rightarrow \infty$.
The $z-$axis sheet conductivity follow from Ohms law,
$\sigma = L \overline{G}_0 /C$,
which in the limit $C, L \rightarrow\infty$ becomes,
\begin{equation}
\sigma = 
{a\over 2b}{t_0^2\over 1-t_0^2}.
\label{sheet_cond}
\end{equation}
Having related the conductivity to the network parameters,
the
mesoscopic crossover lengths $L_0$ and $\xi$ for a {\it finite}
size network model can be readily obtained from
Eqn.~(\ref{L0def}) and (\ref{xidef}).

The exact result for the conductance Eqn.~(\ref{G0}) in the absence of
interference between winding paths should be valid
even for finite circumference, provided $C$ is large enough
so that winding paths are rare.
The  condition for the validity
of ignoring the interference between winding paths
is that $L \ll  \sigma C \sim \xi$,
so that the sample is in the 2d chiral metal or 1d metal regimes.

Notice that $\overline{G}_0$ in Eqn.~(\ref{G0}) is well defined
even as $t_0 \rightarrow 1$.  In this limit, the motion along the z-axis
also becomes ballistic (for finite $N$), and each channel is perfectly transmitted
with $\overline{G}_0 \rightarrow N_c$.  It will be convenient to define
an Ohmic conductance,
\begin{equation}
G_{ohm} \equiv C\sigma/L = \xi/2L  ,
\end{equation}
which coincides with $\overline{G}_0$ when $L$ is large enough
that the $z-$axis motion is diffusive.  As defined,
$G_{ohm}$ diverges with $\sigma$ as $t_0 \rightarrow 1$.
The crossover from diffusive to ballistic $z-$axis motion occurs when
$G_{ohm} \approx N_c$.

The 2d chiral metal regime requires that
$L \ll  L_0$, or equivalently $N \ll  G_{ohm}$.
However, to avoid a crossover into the ballistic
regime of the network model
requires $G_{ohm} < N_c$.  Thus
2d chiral metal behavior is expected
for $N \ll  N_c$.
Since this limit
is difficult to access numerically, we focus below primarily
on the 1d metallic and localized regimes.

\subsection{Results}

\begin{figure}
\epsfxsize=3.0in
\centerline{\epsfbox{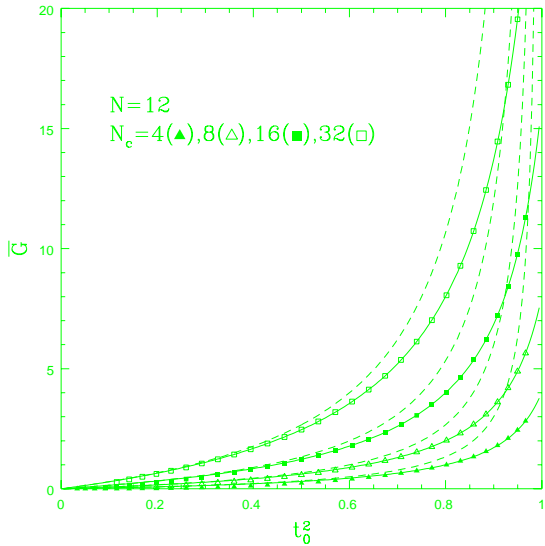}}
\begin{minipage}[t]{8.1cm}
\caption{The mean $z-$axis conductance $\overline{G}$ with fixed height
$N=12$ for several different circumferences, plotted
versus the dimensionless interlayer tunneling probability $t_0^2$.
The solid lines are $\overline{G_0}$ given by Eqn.~(3.6),
and the dashed lines are the Ohmic conductances $G_{ohm}=\xi/2L$
given in Eqn.'s~(3.7) and (3.8).
\label{gt0}}
\end{minipage}
\end{figure}

\begin{figure}
\epsfxsize=3.0in
\centerline{\epsfbox{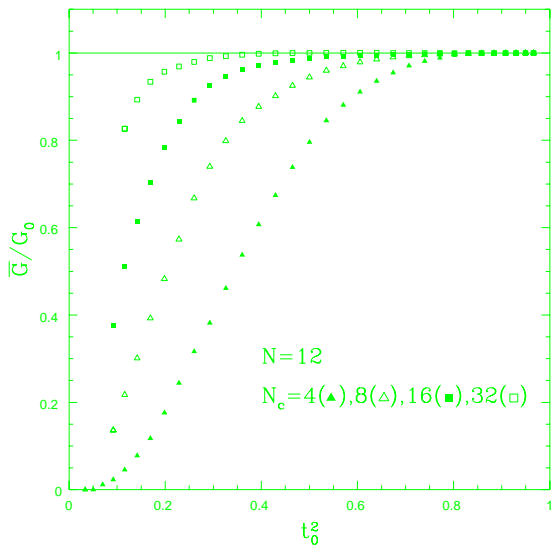}}
\begin{minipage}[t]{8.1cm}
\caption{ The mean conductance $\overline{G}$ from Fig.~5
%\ref{gt0} 
replotted
after normalizing by $G_0$. The ratio $\overline{G}/G_0$ deviates 
from one as the system becomes localized.
\label{ggL}}
\end{minipage}
\end{figure}

In Fig.~\ref{gt0} we show results for the ensemble-averaged
two-terminal conductance, $\overline{G}$,  computed numerically
from the network model, plotted versus
the tunneling parameter $t_0^2$
for various different channel numbers, $N_c$,
at fixed height, $N=12$.  The solid lines are the ``classical"
conductance, $\overline{G}_0$ Eqn.~(\ref{G0}), 
valid in the absence of interference between winding paths, 
and the dashed lines the ``Ohmic conductance", $G_{ohm} = C\sigma/L$.
Notice that $\overline{G}_0$ gives a good fit to the numerical data,
except in the low conductance regime, $\overline{G} < 1$,
where 1d localization effects are expected.  The deviations from 
the classical behavior in this regime can be seen
more easily in Fig.~\ref{ggL}, where we plot the same data
for the conductance, but now normalized by $\overline{G}_0$.  
Strong deviations are seen for small $t_0^2$, where the system
enters into the 1d localized regime and interference between
winding paths is critical. 

\begin{figure}
\epsfxsize=3.0in
\centerline{\epsfbox{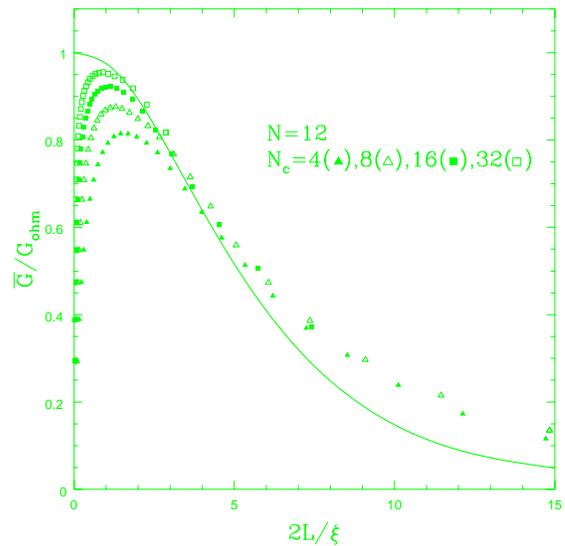}}
\begin{minipage}[t]{8.1cm}
\caption{The mean conductance normalized 
by $G_{ohm}=\xi/2L$ versus $2L/\xi$.
For each sample size the points correspond to different
values of the hopping probability $t_0^2$.
The solid line is the mean conductance computed analytically 
for a quasi-1d system taken from Mirlin {\sl et~al.}'s paper.
\label{gg}}
\end{minipage}
\end{figure}

In order to study the crossover from
the 1d metallic to localized regime,
we plot in Fig.~\ref{gg} the mean conductance for $N=12$, 
normalized by $G_{ohm} = 2\xi/L$,
versus $2L/\xi$.  
The data shows a crossover from a 1d metallic regime with
Ohmic behavior, $\overline{G} \approx G_{ohm}$,
to a 1d localized regime where the conductance
vanishes exponentially for $L \gg  \xi$.
The solid line
is the prediction from Mirlin {\sl et~al.},\cite{mirlin94} 
for the mean conductance of a quasi-1d metallic wire
obtained using supersymmetry methods.
The agreement is reasonable, but our numerics deviate
from the universal form of Mirlin {\sl et~al.}\cite{mirlin94}
both at large and small $L/\xi$.  The deviations at
large $L/\xi$ are presumably due to lattice cutoff
effects,
since in this regime the
localization length along the $z-$axis is
comparable to the network model lattice spacing $a$.
The deviations for small $L/\xi$
are probably due to finite size effects.
Indeed, as the channel number $N_c$ increases,
the agreement improves.
Notice that $\overline{G}/G_{ohm}$ vanishes
as $L/\xi \rightarrow 0$ (rather than approaching unity)
due to ballistic behavior in the network model:
In this limit $t_0 \rightarrow 1$
and
$G_{ohm}$ diverges whereas $\overline{G}$ saturates
at the (finite) channel number $N_c$.  

\begin{figure}
\epsfxsize=3.0in
\centerline{\epsfbox{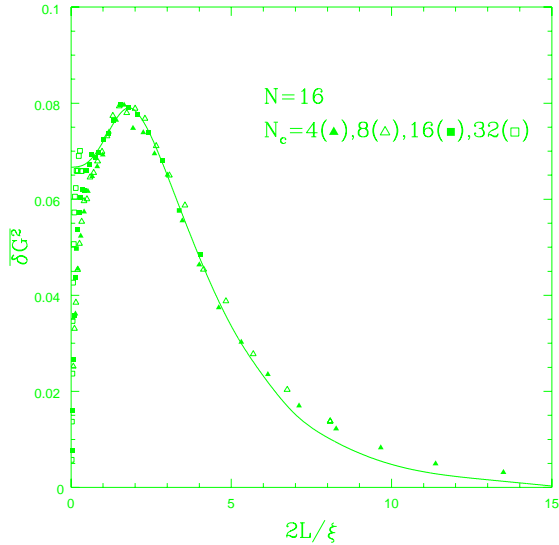}}
\begin{minipage}[t]{8.1cm}
\caption{Variance of the conductance
for different sample sizes and hopping amplitudes
$t_0^2$, all plotted versus $2L/\xi$.
The solid line is the variance of the conductance
for a quasi-1d sample, computed analytically by
Mirlin {\sl et~al.}.
\label{gdev}}
\end{minipage}
\end{figure}

In addition to the mean conductance, we have computed
the sample-to-sample conductance fluctuations.
In Fig.~\ref{gdev} we have plotted $\overline{\delta G^2}$
versus $2L/\xi$, for height $N=16$ and various different
channel numbers.  The solid curve is the universal prediction
for the variance of the conductance of a quasi-1d wire,
obtained by Mirlin {\sl et~al.}.\cite{mirlin94}  This curve shows the crossover
from the 1d metallic regime at small $L/\xi$, where
the variance approaches the well known universal
value, $\overline{\delta G^2} = 1/15$, to the 1d localized
regime where the fluctuations vanish exponentially for $L \gg  \xi$.
The agreement between our numerical data 
and the Mirlin {\sl et~al.}\cite{mirlin94}
theory is quite striking.  Again, the deviations
for $L/\xi \rightarrow 0$ are due to the ballistic
regime in the network model for $t_0 \rightarrow 1$ (with
finite $N$), 
where the conductance fluctuations vanish.
For $L \gg  \xi$ the localization length
approaches the lattice spacing.  The numerics
and theory agree very well
near the peak in the crossover regime.

\begin{figure}
\epsfxsize=3.0in
\centerline{\epsfbox{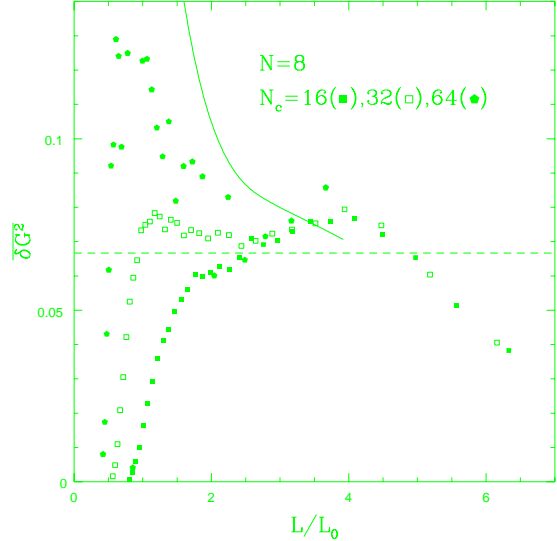}}
\begin{minipage}[t]{8.1cm}
\caption{Variance of the conductance versus $L/L_0$
for three different ``short'' and ``wide'' samples at various values
of $t_0^2$.  The solid line is the conductance variance 
computed analytically by Gruzberg {\sl et~al.}  in the universal 
crossover regime between the 1d and 2d chiral metals.
The dashed line is at $\overline{\delta G^2}=1/15$ - the
value in the 1d metal regime.
\label{gdevL}}
\end{minipage}
\end{figure}

Finally, we mention briefly our effort to extract
numerically the conductance
in the 2d chiral metal regime.  
This regime requires that $L \ll  L_0$,
or equivalently $N \ll  G_{ohm}$.
However, to avoid the ballistic regime when $t_0 \rightarrow 1$,
we must require that $G_{ohm} < N_c$, so that we need
$N \ll  N_c$.  We have focussed on the conductance fluctuations
in this regime, since these are predicted to behave
very differently than in the 1d metal, 
diverging with $L/L_0 \rightarrow 0$ as
$\overline{\delta G^2} \sim (L_0/L)^2$.    
In Fig.~\ref{gdevL} the variance of the conductance is shown
for  ``short" and ``wide" samples,
with height $N=8$ and width
$N_c = 16,32,64$,
plotted versus $L/L_0$ where
$L_0 = \sqrt{a \sigma C}$.  For each width,
$N_c$, we have varied
the tunneling probability, $t_0^2$, to get the set of data points.
The solid line is the analytic prediction from Gruzberg {\sl et~al.},
\cite{gruzberg97}
for the conductance variance in the crossover regime between
the 1d and 2d chiral metal. 
Unfortunately, the agreement with the analytic result is quite poor,
although the agreement improves
for the widest sample with $N_c =64$.
Indeed, the large enhancement in the variance for
the sample with $N_c=64$ in the range $1 < L/L_0 <3$ is
consistent with the theoretical
expectations.
The sharp drop in the conductance
fluctuations for smaller $L/L_0$
is due to the crossover from diffusive
to ballistic motion in the network model.
The local maxima for $N_c = 16$ at $L/L_0\approx 4$ is the same maxima 
as in Fig.~\ref{gdev}, and indicates a crossover into the 1d localized regime
for larger $L/L_0$, where
the Gruzberg {\sl et~al.} results do not apply.  

%%%%%%%%%%%%%%%%%%%%%%%%%%%%%%%%%%%%%%%%%%%%%%%%%%%%%%%%%%%%%%%%%%%%%%%
\section{Inelastic Effects}
%%%%%%%%%%%%%%%%%%%%%%%%%%%%%%%%%%%%%%%%%%%%%%%%%%%%%%%%%%%%%%%%%%%%%%%

The above results for the phase coherent transport are dramatically
modified in the presence of phase breaking effects.  De-phasing effects
can be characterized by a phase breaking time, denoted $\tau_\phi$,
which is the time an electron can propagate before having its phase
randomized by interactions with other electrons or phonons.  In the
extreme anisotropic limit of the surface sheath with vanishing
interlayer tunneling, $t=0$, an electron propagating in one edge
state will interact via Coulomb forces with electrons in neighboring
edges states, and can suffer phase breaking inelastic scattering
events.  Being in 1d, the scattering rate, evaluated to leading order
in the interactions strength $u$, is linear in temperature:
$1/\tau_\phi = c (u l_B/2\pi\hbar v )^2 k_BT/\hbar$, with $c$ an order
one constant, $l_B$ the magnetic length and $v$ the edge velocity.  In
practice, the dimensionless ratio $ul_B/\hbar v$ is itself also of order
one, so that $1/\tau_\phi \sim k_B T/\hbar$.
For non-zero but small interlayer tunneling, the de-phasing rate will
probably crossover to a two-dimensional $T^2$ dependence at very low
temperatures.

Associated with the de-phasing time are two de-phasing lengths:
(i) $l_\phi = v \tau_\phi$, the distance an electron propagates
in the ballistic $x-$direction before de-phasing and (ii)
$L_\phi = \sqrt{D \tau_\phi} = \sqrt{\sigma a l_\phi}$ the
distance an electron diffuses parallel to the field in time $\tau_\phi$.
Consider the transport geometry in Fig.~\ref{system}, 
in which metallic contacts
are applied at $z=0$ and $z=L$.  For $L_\phi \gg L$, an electron
diffuses between the two contacts before being de-phased.  In this
case, transport is {\sl mesoscopic}, and the above phase-coherent
results apply.  

For $L_\phi \ll L$, however, inelastic scattering occurs within the
sample, and we must reconsider transport properties.  There are two
such important {\sl incoherent} regimes, depending upon the relative
magnitude of $l_\phi$ and $C$.  For $l_\phi \ll C$, the electron does
{\it not} fully circumnavigate the sample before suffering a phase
breaking collision.  In this situation, electron paths which wind a
different number of times around the sample do not interfere.  As a
result the system cannot explore the three phase {\it coherent}
regimes discussed in Sections II and III.  Instead, the system is
appropriately described as a phase incoherent 2d chiral metal.
Nevertheless, there are (small) mesoscopic fluctuations expected even
in this limit, which we discuss below.  In the opposite extreme of
$\l_\phi \gg C$, the electron can propagate many times around the
sample before phase breaking.  In this case, the one-dimensional
motion parallel to the field is phase coherent up to a length scale
$L_\phi$.  The system should behave like an incoherent quasi-1d wire,
with $L_\phi$ the appropriate (1d) de-phasing length, as we discuss
further below.

To describe the transport behavior in these incoherent regimes, we
employ arguments first applied in Ref.~\onlinecite{miller60}.  
The important observation is that the sample can be subdivided into
``patches'', whose size is the maximum area over which an electron
diffuses in time $\tau_\phi$.  Each such region effectively acts as a
classical resistor, and the whole sample then as a random resistor
network, the properties of which are well understood.  

First consider $l_\phi \ll C$.  Then the patches have dimensions
$l_\phi$ by $L_\phi$, and form an array of size $C/l_\phi$ by
$L/L_\phi$.  Denoting by $g_i$ the (dimensionless) conductance (along
the $z-$axis) of the $i^{th}$ patch, Ohm's law gives an average patch
conductance of $\overline{g}_i = g_0 = \sigma l_\phi/L_\phi$.  The
conductance fluctuations in each patch, $\delta g_i = g_i - g_0$, are
of order one, - being equivalent to the conductance fluctuations of a
fully coherent network at the boundary between the 1d and 2d metal
regimes.  Since the mean conductance can be written $g_0 = L_\phi/a$,
provided the patch size is larger than the lattice spacing, $L_\phi
\gg a$, the conductance fluctuations in each patch are much smaller
than the mean conductance: $\delta g_i \ll g_0$.  In this limit, both
the total conductance, $\overline{G}$, and it's variance, $\delta G^2
= \overline{G^2} - (\overline{G})^2$, can be easily evaluated.  A
simple estimate is to imagine connecting the resistors (patches) only
vertically (an approximation which gives the correct
result for the conductance fluctuations up to an order one
prefactor).  Then for each column, the patch resistances add, so that
$\delta G_{col}^2 \approx (L_\phi/L)^{3}$, which is independent of
$g_0$.  Contributing in parallel, the conductances of the $N_{col} =
C/l_\phi$ columns add, so that the variance of the {\it total}
conductance is simply $\delta G^2 = N_{col} \delta G_{col}^2$.  This
can be written in the form:
\begin{equation}
  { \delta G^2 \over \overline{G}^2 }
  \approx {a^2 \over {CL}} \left[ {1 \over \sigma}  {l_\phi \over a}
  \right]^{1/2}, 
  \label{fluct_amp}
\end{equation}
with $\overline{G} = C \sigma /L$.  Notice that the conductance
fluctuations have an appreciable temperature dependence entering
through $l_\phi$, growing in magnitude at low temperatures.  The mean
conductance, however, remains temperature independent.

Consider next the 1d incoherent limit with $l_\phi \gg C$,
in which the electron propagates many times around the sample before
de-phasing.  In this limit, the $L/L_\phi$ classical patch resistors form a
one-dimensional random chain, and have dimensions $C$ by $L_\phi$.  
Due to 1d localization effects, the conductance of each such segment
will depend strongly on it's length, $L_\phi$, and hence on the temperature
$T$.  For example, when $L_\phi$ is much smaller than 
the 1d localization length $\xi$, the (mean) conductance
of each segment is given by,
\begin{equation}
  G_{seg}(L_\phi) = (\sigma C/L_\phi) - {2 \over 45} {L_\phi \over \xi}
  +O(L_\phi/\xi)^2    ,
\end{equation}
where the first term is Ohm's law, and the second term reflects
the leading 1d localization corrections within the unitary ensemble.
In the opposite limit, $L_\phi \gg \xi$, one expects
a stronger length (and temperature) dependence, $G_{\rm seg}(L_\phi) \sim 
\exp(-L_\phi/2\xi)$.  The total conductance follows by simply adding
the series resistances of each of the $L/L_\phi$ segments.
In the 1d metallic regime with $L_\phi \ll \xi$, this gives,
\begin{equation}
\overline{G} = { \sigma C \over L} - {2 \over 45} {L_\phi \over L}
 {L_\phi \over \xi} + ...   ,
\end{equation}
which depends on temperature through $L_\phi(T)$.  

%As one cools the sample and $l_\phi$ increases, the conductance will
%start showing a temperature dependence (of the above form) when
%$l_\phi > C$.  More precisely, the 1d localization corrections
%responsible for the temperature dependence require the interference
%between two returning paths, both which twice loop back to the origin
%on the same two loops, but taken in different order.  (This is
%analogous to the interference between time reversed paths in
%conventional weak localization.)  For the chiral surface sheath, this
%requires that an electron can fully circumnavigate the sample {\it
%  twice} without suffering de-phasing collisions.

Experimentally, such conductance fluctuations are usually observed not
by looking at different samples, but by varying the applied magnetic
field in such a way as to change the phases accumulated by interfering
electrons and thereby effectively change the disorder.  The
conductance fluctuations in this context are characterized not only by
their amplitude, discussed above, but also by a characteristic field
scale $\delta B_\phi$.  This scale is defined by the amount the
applied field must be changed in order that the conductance of a fixed
sample becomes uncorrelated with its previous value.  Physically, the
conductance fluctuations arise from constructive interference of two
paths enclosing an area of the phase-coherent patch size.  The total
change in phase shift around this loop in units of $2\pi$ is simply
the change in magnetic flux through this area divided by the flux
quantum $\phi_0 = hc/e$.  The characteristic field $\delta B_\phi$,
which changes the phase around the loop by $O(\pi)$, is thus simply
the field which puts, say, half a flux quantum through this coherent
area.  Assuming the magnetic field has a non-negligible angle to the
surface sheath (which we believe to be the case in the experiments of
Druist {\sl et~al.}), this gives
\begin{equation}
  \delta B_\phi \approx \cases{ \phi_0/l_\phi L_\phi & $l_\phi \ll C$
    \cr
    \phi_0/ C L_\phi & $l_\phi \gg C$ \cr},
  \label{dB_scale}
\end{equation}
in the two incoherent regimes.  Note that since $l_\phi$ and $L_\phi$
increase as temperature is lowered, the conductance varies very
rapidly with field at low temperatures.

\section{Conclusions}

We conclude with a comparison of these theoretical results to the
experimental data of Druist {\sl et~al.}\cite{druist97}.  
Druist {\sl et al.} have measured the z-axis transport
in a series of multilayer quantum Hall samples.
Specifically, the samples consisted of 50 layers of
$150 \AA$ $GaAs$ layers alternating with
$150 \AA$ $Al_{0.1}Ga_{0.9}As$ barriers doped at their centers
with Silicon.  The vertical separation between each of the 50 2d electron gases
is $a=300 \AA$.  A simple Kronig-Penney analysis
gives an estimate for the z-axis bandwidth of $t = 0.12meV$.
When the applied magnetic field was tuned onto an
integer quantum Hall plateau, the z-axis conductance -  dropping with
temperature -  was found to saturate below about $200mK$.
Since the low temperature z-axis conductance scaled
linearly with the circumference (perimeter) of the samples,
which ranged from $400\mu m \le C \le 7mm$,
Druist et. al. argued that the conduction was being dominated
by the 2d chiral surface sheath.
The resulting sheet {\it conductivity} along the z-axis was
found to be $\sigma \approx 4 \times 10^{-4}$ on the $\nu=1$ plateau,
and about a factor of three larger for $\nu=2$.

A theoretical estimate for the z-axis conductivity
of the surface sheath at one full Landau level can be obtained from\cite{balents96}
\begin{equation}
  \sigma \approx {a l_0 \over\hbar^2v^2}t^2,
\end{equation}
where $l_0$ is an elastic mean free path for edge scattering
and $v$ is the (ballistic) edge velocity.
Unfortunately, both $v$ and $l_0$ are difficult to
estimate reliably, depending on the detailed slope and irregularities
of the edge
confining potential.    
However, we expect that in the
limit of large magnetic field, $l_0 \gtrsim l_B$, where $l_B$
is the magnetic length
($l_0$ may grow much longer than $l_B$
as the edge is made cleaner).  
Moreover, we expect $v$ to be bounded above
by the edge velocity for a hard-wall confining potential,
so that $v
\lesssim \omega_c l_B/2\pi$, with $\omega_c$ the cyclotron
frequency.
Putting in these (rough) bounds, we obtain
\begin{equation}
  \sigma \gtrsim  {{(2\pi)^2 t^2 a} \over {\hbar^2 \omega_c^2 l_B}}.
\end{equation}
Using the parameters appropriate for the Druist {\sl et~al.}
experiment, this gives $\sigma \gtrsim 6\times 10^{-5}$, about an
order of magnitude smaller than the experimental value.  Given the
uncertainties in $v$ and $l_0$, as well as possible shifts in $t$ due
to interaction effects, this level of agreement is reasonable.

Taking now the {\sl measured} value of $\sigma$, we can estimate
the two length scales which determine the system behavior in the
mesoscopic limit.  
The samples studied by Druist {\sl et~al.} had
a range of circumferences $400 \mu m \leq C \leq 7 mm$,
which  
correspond to lengths $2 \leq L_0/a \leq 10$ and $10 \leq \xi/a \leq
200$, upon using Eqns.~(\ref{L0def})--(\ref{xidef}).  Since $N=L/a = 50$ in these
experiments, in the {\sl mesoscopic} limit these samples should span
the quasi-1d metal and 1d localized regimes.  At low temperatures, we
would therefore expect a strong suppression of the conductivity and
significant temperature and circumference dependence, especially in
the smaller samples.
That such effects are not observed must be attributed to {\it inelastic}
effects.  
Indeed, as shown below,
estimates for the in-plane de-phasing length
$l_\phi$ give $l_\phi \ll C$ even at the lowest temperatures
and for the smallest sample.  
In this limit, mesoscopic effects are greatly suppressed,
and the system is best thought of as an {\it incoherent} 2d chiral metal.
This accounts naturally for the observed low temperature
saturation of the conductivity (it remains to be
seen whether the weak residual temperature dependence at low $T$ can
be fitted to the expected\cite{balents96}\ form $\sigma(T) -
\sigma(0) \propto T^2$).  

We can attempt to estimate the de-phasing length $l_\phi$ via
\begin{equation}
  l_\phi = A \left( {{h v} \over {u l_B}} \right)^2 {{hv} \over
    {k_{\rm B}T}},
\end{equation}
however there is considerable uncertainty in the parameters - 
particularly the edge velocity $v$.  As a crude estimate
we take $A=1$, a dimensionless interaction strength of unity
$ul_B/hv =1$ and an edge velocity estimated for a hard-wall
confining potential $v = \omega_cl_B/2\pi$.  In the $10$ Tesla
field used by Druist {\sl et~al.} in the $\nu=1$ plateau
and at the lowest temperatures studied of 
$T=50 mK$ this gives the rough estimate
$l_\phi \sim 20\mu m$
  
Fortunately, one can also extract estimates for $l_\phi$ 
directly from the
experimentally measured conductance fluctuations.  In fact, this can
be done in two ways, thereby providing a consistency check.  One
determination is from the {\sl amplitude} of the fluctuations.  Solving
Eqn.~(\ref{fluct_amp}) gives
\begin{equation}
  l_\phi \approx \tilde{A} {C N^3 \over \overline{G}^3} (\delta G^2)^2.
\end{equation}
Because the fourth power of $\delta G$ appears above and the
amplitude $\tilde{A}$ is unknown, there is again considerable
uncertainty in $l_\phi$.  For the Druist {\sl et~al.}  experiments,
we obtain $l_\phi \approx 26 \mu m$, consistent
with the above theoretical estimate.

A second determination comes from the magnetic field scale of the
conductance fluctuations.  From the above estimates, we see that
$L_\phi = \sqrt{\sigma a l_\phi} \lesssim a$ (using the measured
$\sigma = 4 \times 10^{-4}$).  This is close to the ``incoherent
tunneling'' limit, and we expect it is appropriate to replace $L_\phi
\rightarrow a$ in Eqn.~(\ref{dB_scale}), giving
\begin{equation}
  l_\phi \approx {\phi_0 \over {a \delta B_\phi}}.
\end{equation}
For the Druist et. al. experiment, this gives $l_\phi \approx 3 \mu m$
at $T=100 mK$, somewhat smaller than the first estimate.  In this case
there are also considerable uncertainties due primarily to incomplete
knowledge of the degree of interlayer flux penetration.  However, all
three of the above estimates give $l_\phi \ll C$.

In summary, the experiments so far are consistent with the picture of
an {\sl incoherent 2d chiral metal}.  Several opportunities exist for
further theoretical and experimental study.  
Samples with smaller circumferences in the range
of $10$ to $20 \mu m$ would
be highly desirable, since the mesoscopic regime
would then be accessible below several hundred $mK$.
In this limit, the rich and varied crossovers between the three mesoscopic regimes could be accessed experimentally.
Theoretically, a more quantitative study of
inelastic scattering and de-phasing lengths would be desirable in order
to achieve a precise comparison with experiment.  Particularly
interesting from both points of view is the temperature dependence of
$1/\tau_\phi$, which we believe should exhibit linear scaling with
temperature over a broad range.  A field-theoretic treatment of
de-phasing effects could be useful in providing the desired tighter
link with experiments.

% All the samples contains fifty quantum Hall layers, but
% it is tested by an in-plane transport measurement 
% that only forty eight layers contribute.
% Thus, the height of the system $L=1.44\mu m$.
%and the areas range between $1\times 10^4\mu m^2$ and $4.22\times
%10^5\mu m^2$.  The measurements are done at 
%a low enough temperature where the conductance
%becomes independent of the temperature $T<100mK$.
%For the following experimental values,

%\begin{eqnarray}
% t & = & 0.05 meV\nonumber\\
% a & = & 300 {\AA}\nonumber\\
% v & \approx & 4\times 10^6 cm/s\nonumber\\
% \tau & \approx & 1\times 10^{-12} s,\nonumber
%\end{eqnarray}

%%%%%%%%%%%%%%%%%%%%%%%%%%%%%%%%%%%%%%%%%%%%%%%%%%%%%%%%%%%%%%%%%%%%%%%
\section*{Acknowledgments}
%%%%%%%%%%%%%%%%%%%%%%%%%%%%%%%%%%%%%%%%%%%%%%%%%%%%%%%%%%%%%%%%%%%%%%%

We thank David Druist and Elizabeth Gwinn for generously sharing 
their experimental data.
It is a pleasure to acknowledge fruitful conversations with 
Ilya A. Gruzberg, Nick Read and Hsiu-Hau Lin.
We are grateful to the National Science Foundation for support, under
Grant Nos. PHY94-07194, DMR-9400142, and DMR-9528578.

\bibliography{/usr/home/spock/sora/paper/biblio/ref}

\end{multicols}
\end{document}